\documentclass[runningheads]{svmult}

\usepackage{makeidx}   
\usepackage{graphicx}  
\usepackage{subeqnar}  
\usepackage{multicol}  
\usepackage{physprbb}  
\makeindex             

\def\sigmat{\sigma_\mathrm{T}}
\def\me{m_\mathrm{e}}

\def\ms{M_\odot}
\def\ledd{L_\mathrm{Edd}}
\def\tc{T_\mathrm{C}}
\def\teq{T_\mathrm{eq}}
\def\mbh{M_\mathrm{BH}}
\def\mcrit{M_\mathrm{\rm BH, crit}}
\def\tvir{T_\mathrm{vir}}
\def\rc{R_\mathrm{C}}
\def\mgas{M_\mathrm{gas}}

\def\mstar{M_*}

\def\ncrit{n_{\rm crit}}

\def\reff{R_{\rm e}}
\def\etaesc{\eta_{\rm esc}}

\def\lesssim{\hbox{\rlap{\hbox{\lower4pt\hbox{$\sim$}}}\hbox{$<$}}}
\def\gtrsim{\mathrel{\hbox{\rlap{\hbox{\lower4pt\hbox{$\sim$}}}\hbox{$>$}}}}
\def\beq{\begin{equation}}
\def\eeq{\end{equation}}
\def\beqa{\begin{eqnarray}}
\def\eeqa{\end{eqnarray}}

\begin{document}

\title*{Radiative Feedback from Quasars and the Growth of Supermassive
Black Holes}

\toctitle{Radiative Feedback from Quasars and the Growth of Supermassive
Black Holes}

\titlerunning{Radiative Feedback from Quasars}

\author{Sergey Yu. Sazonov\inst{1,2}
\and Jeremiah P. Ostriker\inst{3}
\and Luca Ciotti\inst{4}
\and Rashid A. Sunyaev\inst{1,2}
}
\authorrunning{Sergey Yu. Sazonov et al.}

\institute{Max-Planck-Institut f\"ur Astrophysik,
Karl-Schwarzschild-Str. 1,\\
85740 Garching, Germany
\and
Space Research Institute, Profsoyuznaya 84/32, 117997 Moscow, Russia
\and
Institute of Astronomy, Madingley Road, CB3 0HA Cambridge
\and
Department of Astronomy, University of Bologna, via Ranzani 1,\\
I-40127 Bologna, Italy
}

\maketitle

\begin{abstract}
We discuss the role of feedback via photoionization and
Compton heating in the co-evolution of massive black holes at
the center of spheroidal galaxies and their stellar and gaseous
components. We first assess the energetics of the radiative feedback
from a typical quasar on the ambient interstellar gas. We then
demonstrate that the observed $\mbh$--$\sigma$ relation could be
established at a relatively early epoch in galactic evolution when the
formation of the stellar bulge was almost completed and the
gas-to-stars mass ratio was reduced to a low level $\sim$0.01 such
that cooling could not keep up with radiative heating. A considerable
amount of gas was expelled at that time and black hole accretion
proceeded at a much lower rate thereafter.  
\end{abstract}

\section{General Picture}
\label{intro}

Most elliptical galaxies are poor with respect to
interstellar gas. Also, elliptical galaxies
invariably contain central massive black holes (BHs), and there exists a tight
relationship between the characteristic stellar velocity dispersion
$\sigma$ and the BH mass $\mbh$ \cite{fm00,tgb+02},
and between $\mbh$ and the host spheroid mass in stars, $\mstar$
\cite{mtr+98}. Are these two facts related? Here 
we focus on a scenario in which the mass of the central BH grows 
within gas rich elliptical progenitors until star formation has
reduced the gas fraction in the central regions to of order 1\% of the
stellar mass. Then radiative feedback during episodes when 
the BH luminosity approaches its Eddington
limit drives much of the gas out of the galaxy, limiting both future
growth of the BH and future star formation to low levels. 

Many works already recognized the importance of feedback as a key
ingredient of the mutual BH and galaxy evolution
\cite{bt95,co97,sr98,f99,co01,bs01,cv01,cv02,k03,wl03,hco04,gds+04,mqt04}.
What is new  about this work is the stress on one component of the problem
that has had relatively little attention: the radiative output of the
central BH is not conjectural -- it must have occured -- and
the high energy component of that radiative output will have a
dramatic and calculable effect in heating the gas in ellipticals.

Using the average quasar spectral output derived in \cite{sos04}, we
show below and in more detail in \cite{soc+04} that the limit on the
central BH produced by the above argument coincides accurately with
the observed $\mbh$--$\sigma$ relation. Not only the slope, but also the small 
normalization factor is nicely reproduced.

The present work is complementary to \cite{co97,co01} in that, while
it does not attempt to model the complex flaring behavior of an
accreting BH with an efficient hydrodynamical code, it does do a far more
precise job of specifying the input spectrum and the detailed atomic
physics required to understand the interaction between that spectrum
and the ambient interstellar gas in elliptical galaxies.

\section{Radiative Heating of ISM in Spheroids}
\label{heatfar}

Below we assess the conditions required for the central BH radiation
to significantly heat the ISM over a substantial volume of the
galaxy. In this section we shall {\sl assume} that the central
BH has a mass as given by the observed $\mbh$--$\sigma$ relation for 
local ellipticals and bulges \cite{tgb+02}:
\beq
\mbh=1.5\times 10^8\ms\left(\frac{\sigma}{200~{\rm km}~{\rm s}^{-1}}\right)^4.
\label{sigma_mbh}
\eeq
This assumption will be dropped in \S\ref{origin}, where
we {\sl predict} the $\mbh$--$\sigma$ relation. 

\begin{figure}
\centering
\includegraphics[bb=10 170 570 680, width=0.68\columnwidth]{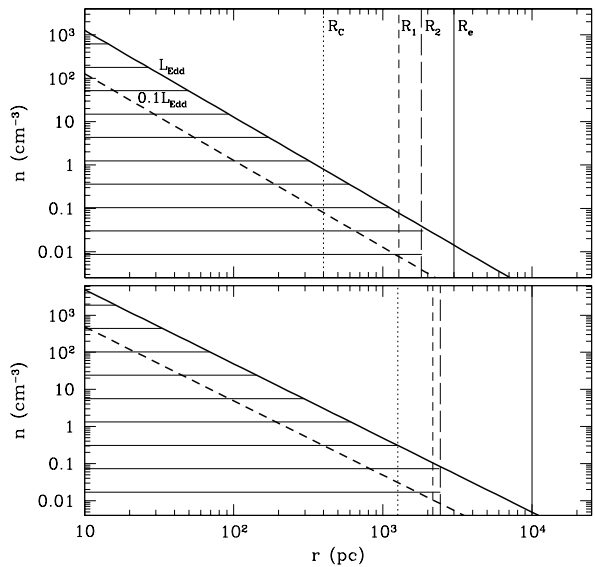}
\caption{The $(r,n)$ plane for a galaxy with
$\sigma=180$\,km\,s$^{-1}$ ($\tvir=2.4\times 10^6$\,K, $\mbh=10^8\ms$, 
upper panel), and with $\sigma=320$\,km\,s$^{-1}$ ($\tvir=7.7\times
10^6$\,K, $\mbh=10^9\ms$, lower panel). In the dashed area, gas can be
heated above $\tvir$ by radiation from the central BH emitting at the
Eddington luminosity. The upper boundary of this area scales linearly
with luminosity. Vertical boundaries are $\rc$, $R_1$, $R_2$ and $\reff$.
} 
\label{nr_tvir}
\end{figure}

In \cite{sos04} we computed for the average quasar spectrum the equilibrium
temperature $\teq$ (at which heating due to Compton scattering and
photoionization balances cooling due to line and continuum emission) of gas of
cosmic chemical composition as a function of the ionization parameter
$\xi\equiv L/nr^2$, where $L$ is the BH bolometric luminosity. In the
range $3\times 10^4$--$10^7$\,K, 
\beq
\teq(\xi)\simeq 2\times 10^2\xi\,{\rm K},
\label{tstat_eq}
\eeq
while at $\xi\ll 10^2$ and $\xi\gg 10^5$, $\teq\approx 10^4$\,K and $2\times
10^7$\,K, respectively. 
On the other hand, the galactic virial temperature is given by
\beq	
\tvir\simeq 3.0\times 10^6\,{\rm K}
\left(\frac{\sigma}{200~{\rm km}~{\rm s}^{-1}}\right)^2.
\label{tvir}
\eeq
We can then find the critical density $\ncrit$ defined by
\beq
\teq(L/\ncrit r^2)=\tvir
\label{ncrit}
\eeq 
as a function of distance $r$ from the BH. Gas with $n<\ncrit(r)$ will be
heated above $\tvir$ and expelled from the galaxy. We show in
Fig.~\ref{nr_tvir} the resulting $(r,n)$ diagrams for a small and
large BH/galaxy.

In reality, provided that $\teq>\tvir$, significant heating  will take
place only out to a certain distance that depends on the luminosity
and duration of the quasar outburst. Since the BH releases via 
accretion a finite total amount of energy, $\epsilon\mbh c^2$, there 
is a characteristic limiting distance:  
\beq
\rc=\left(\frac{\sigmat\epsilon\mbh}{3\pi\me}\right)^{1/2}
     = 400\,{\rm pc}\left(\frac{\epsilon}{0.1}\right)^{1/2}
          \left(\frac{\mbh}{10^8\ms}\right)^{1/2}.
\label{rc}
\eeq
Inside this radius, a low density, fully photoionized gas will be heated to
the Compton temperature $\tc\approx 2$\,keV characteristic of the
quasar spectral output. More relevant for the problem at hand is
the distance out to which low density gas will be Compton heated to
$T\gtrsim\tvir$:  
\beq
R_1=\rc\left(\frac{\tc}{\tvir}\right)^{1/2}=
    1,300\,{\rm pc}\left(\frac{\epsilon}{0.1}\right)^{1/2}
       \frac{\sigma}{200~{\rm km}~{\rm s}^{-1}}.
\label{r1}
\eeq
Yet another characteristic radius is out to which gas of critical
density $\ncrit$ will be heated to $T\gtrsim\tvir$ via photoinization
and Compton scattering:
\beq R_2=R_1 [\Gamma(n_{\rm crit})/\Gamma_{\rm C}]^{1/2},
\label{r2}
\eeq
where $\Gamma_{\rm C}$ and $\Gamma$ are the Compton and total heating rates,
respectively. Depending on gas density ($0<n<n_{\rm crit}$), the outer
boundary of the ``blowout region'' will be located somewhere between $R_1$ and
$R_2$. 
The size of the heating zone can be compared with the galaxy effective
radius
\beq
\reff\sim 4,000\,{\rm pc}
\left(\frac{\sigma}{200~{\rm km}~{\rm s}^{-1}}\right)^2.
\label{reff}
\eeq

\begin{figure}
\centering
\includegraphics[bb=10 170 570 680, width=0.65\columnwidth]{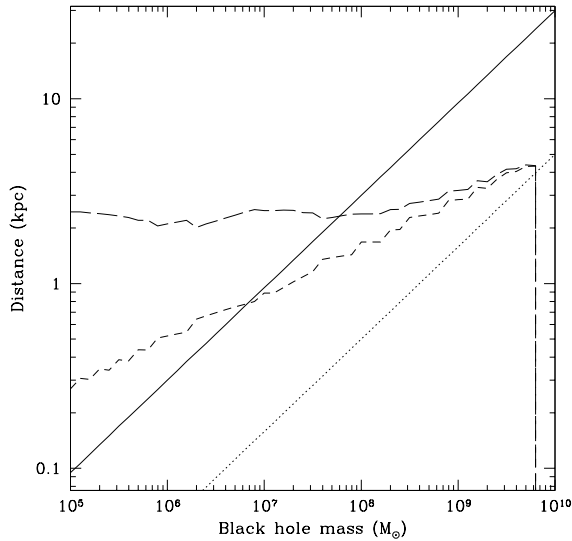}
\caption{Different heating radii: $\rc$ (dotted line),
$R_1$ (short-dashed line), and $R_2$ (long-dashed line), and the galactic
effective radius (solid line), as a function of $\mbh$. 
}
\label{mbh_radii}
\end{figure}

The different characteristic distances defined above are shown as a
function of $\mbh$ in Fig.~\ref{mbh_radii}. One can see that a BH of
mass $<10^7\ms$ should be able to unbind the ISM out to several
$\reff$. In the case of more massive BHs/galaxies with $\mbh\sim
10^8$--$10^9\ms$, the heating will be localized to innermost
$\sim$0.3--0.5$\reff$.

\section{Possible Origin of the $\mbh$--$\sigma$ Relation}
\label{origin}

We now consider the following general idea. Before the BH grows to a
certain critical mass, $\mcrit$, its radiation will be unable to
efficiently heat the ambient gas, and the BH will accrete gas
efficiently. Once the BH has grown to $\mcrit$, its radiation will
heat and expell a substantial amount of gas from the central
regions. Feeding of the BH will then become self-regulated on a time
scale of order the cooling time of the low density gas. Subsequent
central activity will be characterized by a very small duty cycle
($\sim$0.001), as predicted by hydrodynamical simulations
\cite{co97,co01} and suggested by observations \cite{hco04}. BH
growth will be essentially terminated.

Suppose that the galaxy density distribution is that of a singular isothermal
sphere, with the gas density following the total density:
\beq
\rho_{\rm gas}(r)=\frac{\mgas}{M}\frac{\sigma^2}{2\pi Gr^2}.
\label{rho_gas}
\eeq
Here $\mgas$ and $M$ are the gas mass and total mass within the
region affected by radiative heating. The size of the
latter is uncertain but is less than a few kpc (see \S\ref{heatfar}),
so that $M$ is dominated by stars rather than by dark matter.

Radiation from the central BH can heat the ambient gas up to 
\beq
\teq\approx 6.5\times 10^3\,{\rm K}
\frac{L}{\ledd}\left(\frac{\mgas}{M}\right)^{-1}\frac{\mbh}{10^8\ms}
               \left(\frac{200\,{\rm km~s}^{-1}}{\sigma}\right)^2,
\label{tstat}
\eeq
this approximate relation being valid in the range $3\times
10^4$--$10^7$\,K. Remarkably, $\teq$ does not depend
on distance for the adopted $r^{-2}$ density distribution. 
We then associate the transition from rapid BH growth to slow,
feedback-limited BH growth with the critical condition  
\beq
\teq=\etaesc\tvir, 
\label{tstat_tvir}
\eeq
where $\etaesc\gtrsim 1$ and $\tvir$ is given by (\ref{tvir}). Once
heated to $\teq\gtrsim\tvir$, the gas will stop feeding the BH. The
condition (\ref{tstat_tvir}) will be met for
\beq
\mcrit=4.6\times 10^{10}\ms\etaesc\left(\frac{\sigma}
       {200\,{\rm km~s}^{-1}}\right)^4\frac{\ledd}{L}\frac{\mgas}{M}.
\label{mcrit}
\eeq 
Therefore, for fixed values of $\etaesc$, $L/\ledd$ and $\mgas/M$ we
expect $\mcrit\propto\sigma^4$, similarly to the observed
$\mbh$--$\sigma$ relation. Equally important is the 
normalization of the $\mbh$--$\sigma$ relation. By comparing
(\ref{mcrit}) with (\ref{sigma_mbh}) we find that the observed
correlation will be established if 
\beq 
\mgas/M=3\times 10^{-3}\etaesc^{-1}L/\ledd.  
\label{mgas_crit}
\eeq 
The gas-to-stars ratio is thus required to be low and approximately
constant for spheroids of different mass at a certain stage of their
evolution. As for the Eddington ratio, it is reasonable to expect
$L/\ledd\sim$0.1--1 during quasar outbursts.

\begin{figure}
\centering
\includegraphics[bb=10 170 570 680, width=0.65\columnwidth]{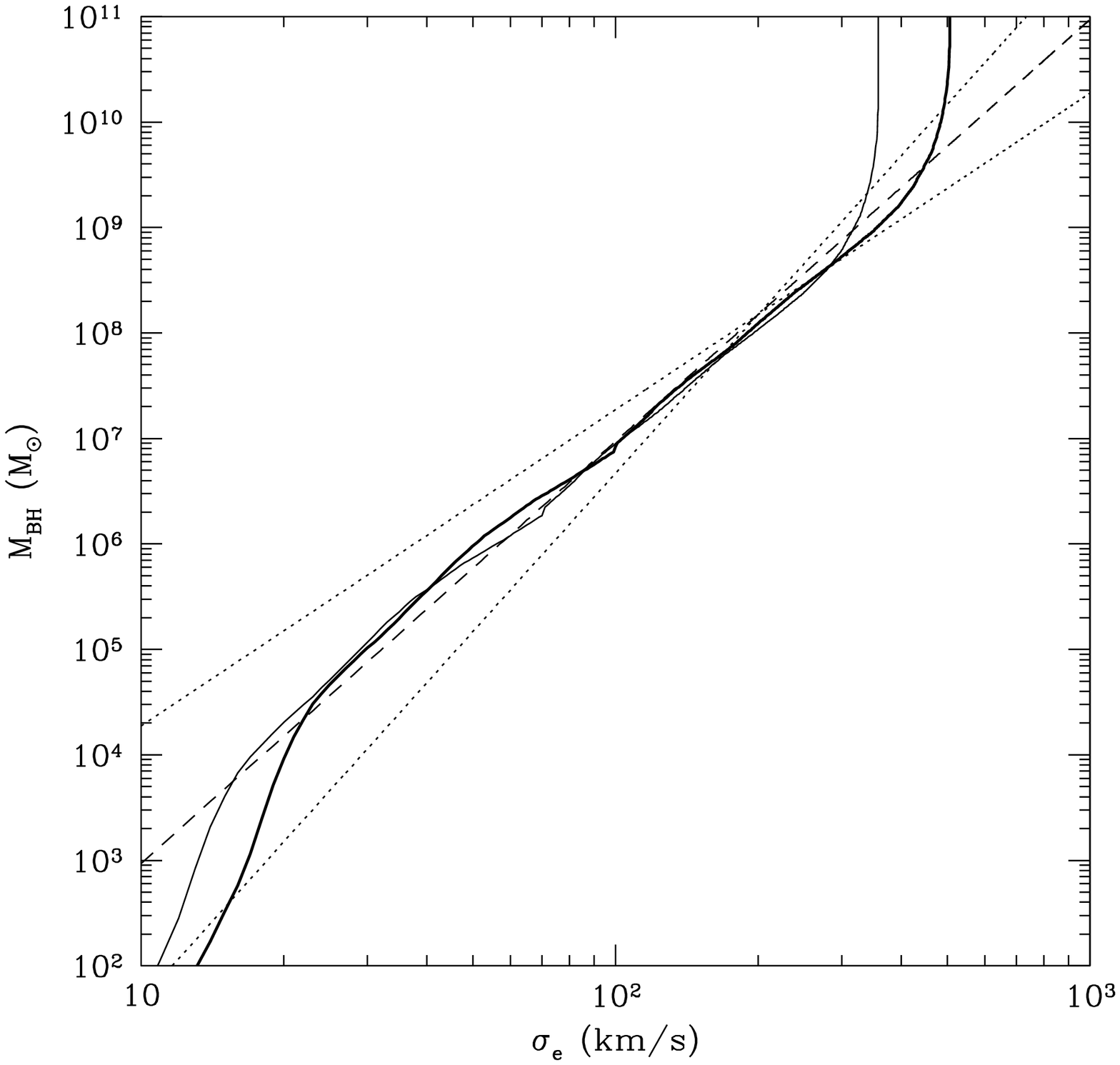}
\caption{Thick solid line shows the predicted $\mbh$--$\sigma$ correlation
resulting from heating of the ISM by the radiation from the central BH
assuming $\mgas(\reff)/M=0.003$ and $\etaesc=1$. Thin solid line
corresponds to $\mgas(\reff)/M=0.0015$ and $\etaesc=2$. Dashed line is the
observed $\mbh\propto\sigma^4$ correlation in the range $10^6$--a few 
$10^9\ms$, extrapolated to lower and higher BH masses. Dotted lines
are $\mbh\propto\sigma^3$ and $\mbh\propto\sigma^5$ laws.  
}
\label{s_mbh}
\end{figure}

The approximately linear $\teq(\xi)$ dependence  [see
(\ref{tstat_eq})] was crucial to the above argument leading 
to the $\mcrit\propto\sigma^4$ result. However, the $\teq(\xi)$
function becomes nonlinear outside the range $3\times 10^4\,{\rm
K}<\teq<10^7\,{\rm K}$ \cite{sos04}. In Fig.~\ref{s_mbh} 
we show the predicted correlation between $\mcrit$ and $\sigma$ for
$L/\ledd=1$ and $\mgas/M=3\times 10^{-3}$. It can be seen that the
$\mbh\propto\sigma^4$ behavior is expected to break down for $\mbh<10^4\ms$ and
for $\mbh\gtrsim$~a few $10^9\ms$. It is perhaps interesting that the
range of masses shown in Fig.~\ref{s_mbh} for which 
$\mbh\propto\sigma^4$ is obtained from considerations of atomic
phyisics (and the observed AGN spectra) corresponds closely with the
range of masses for which this power law provides a good fit to the
observations. Exploring the $\mbh$--$\sigma$ relation observationally
near $10^9\ms$ would be a sensitive test of the importance of
radiative feedback.

\section{Detailed Modelling of the BH-Galaxy Co-evolution}

In \cite{soc+04,cos+04,oc04} we addressed in a more
quantitative way the BH growth in the context of the parent galaxy
evolution. We adopted a physically-motivated one-zone model, taking
into account the mass and energy return from the evolving stellar
population. This model predicts that after an initial ``cold'' phase
dominated by gas infall, once the gas density becomes sufficiently low
the gas heating dominates and the galaxy switches to a ``hot''
solution. The gas mass/stellar mass ratio at that epoch
($\sim$0.003) is remarkably close to the value inferred above from the
argument leading to the right $\mbh$--$\sigma$ relation. Other
predictions of the toy model are also in satisfactory agreement with
observations. The ``cold'' phase would probably be identified
observationally with the Lyman Break and SCUBA galaxies, while the
``hot'' phase with normal, local ellipticals. 

A proper investigation of the importance of radiative heating on the
BH/galaxy co-evolution, based on hydrodynamical numerical simulations,
is now in progress.


%

\end{document}